\documentclass[english,superscriptaddress,showpacs,showkeys]{revtex4-2}
\usepackage[T1]{fontenc}
\usepackage[utf8]{inputenc}
\usepackage{color}
\usepackage{babel}
\usepackage{units}
\usepackage{amsmath}
\usepackage{amssymb}
\usepackage{graphicx}
\usepackage[pdfusetitle,
 bookmarks=true,bookmarksnumbered=false,bookmarksopen=false,
 breaklinks=false,pdfborder={0 0 0},pdfborderstyle={},backref=false,colorlinks=true]
 {hyperref}
\hypersetup{
 citecolor=blue,linkcolor=blue,urlcolor=blue}
\begin{document}
\title{Natural explanation of recent results on $e^{+}e^{-}\to\Lambda\bar{\Lambda}$}
\author{A.I. Milstein}
\email{A.I.Milstein@inp.nsk.su}

\author{S.G. Salnikov}
\email{S.G.Salnikov@inp.nsk.su}

\affiliation{Budker Institute of Nuclear Physics of SB RAS, 630090 Novosibirsk,
Russia}
\affiliation{Novosibirsk State University, 630090 Novosibirsk, Russia}
\date{\today}
\begin{abstract}
We show that the recent experimental data on the cross section of
the process $e^{+}e^{-}\to\Lambda\bar{\Lambda}$ near the threshold
can be perfectly explained by the final-state interaction of $\Lambda$
and $\bar{\Lambda}$. The enhancement of the cross section is related
to the existence of low-energy real or virtual state in the corresponding
potential. We present a simple analytical formula that fits the experimental
data very well.
\end{abstract}
\maketitle
Recently, new experimental data have appeared on the cross section
of $e^{+}e^{-}\to\Lambda\bar{\Lambda}$ annihilation near the threshold~\citep{Ablikim2023}.
These data are consistent with the results of previous works~\citep{Aubert2007,Ablikim2018,Ablikim2019c},
but have much higher accuracy. All these results demonstrate a strong
energy dependence of the cross section near the threshold. A similar
phenomenon has been observed in such processes as $e^{+}e^{-}\to p\bar{p}$~\citep{Aubert2006,Lees2013,Akhmetshin2016,Ablikim2020,Ablikim2021b,Akhmetshin2019,Ablikim2015,Ablikim2019},
$e^{+}e^{-}\to n\bar{n}$~\citep{Achasov2014,Ablikim2021f,Achasov2022},
$e^{+}e^{-}\to\Lambda_{c}\bar{\Lambda}_{c}$~\citep{Pakhlova2008,Ablikim2018b},
$e^{+}e^{-}\to B\bar{B}$~\citep{Aubert2009}, and others. In all
these cases the shapes of near-threshold resonances differ significantly
from the standard Breit-Wigner parameterization. The origin of the
phenomenon is naturally explained by the strong interaction of produced
particles near the threshold (the so-called final-state interaction).
Since a typical value of the corresponding potential is rather large
(hundreds of MeV), existence of either low-energy bound state or virtual
state is possible. In the latter case a small deepening of the potential
well leads to appearance of a real low-energy bound state. In both
cases, the value of the wave function (or its derivative) inside the
potential well significantly exceeds the value of the wave function
without the final-state interaction. As a result, the energy dependence
of the wave function inside the potential well is very strong. Since
quarks in $e^{+}e^{-}$ annihilation are produced at small distances
of the order of $1/\sqrt{s}$, a strong energy dependence of the cross
section is determined solely by the energy dependence of the wave
function of produced pair of hadrons at small distances. Such a natural
approach made it possible to describe well the energy dependence of
almost all known near-threshold resonances (see Refs.~\citep{Haidenbauer2014,Haidenbauer2016,Haidenbauer2021,Milstein2021,Milstein2022a,Milstein2022,Milstein2022c}
and references therein).

The annihilation $e^{+}e^{-}\to\Lambda\bar{\Lambda}$ near the threshold
is the most simple for investigation. This is due to the fact that
the $\Lambda\bar{\Lambda}$ system has a fixed isotopic spin $I=0$,
and the pair is produced mainly in the state with an angular momentum
$l=0$ (the contribution of state with $l=2$ can be neglected). Moreover,
there is no Coulomb interaction between $\Lambda$ and $\bar{\Lambda}$.
Our analysis shows that the imaginary part of the optical potential
of $\Lambda\bar{\Lambda}$ interaction, which takes into account the
possibility of annihilation of $\Lambda\bar{\Lambda}$ pair into mesons,
has only little effect on the cross section. Therefore, we neglect
the imaginary part of the potential. Finally we describe the cross
section of the process $e^{+}e^{-}\to\Lambda\bar{\Lambda}$ by a simple
analytical formula (see Ref.~\citep{Milstein2022c} and references
therein for more details):
\begin{equation}
\sigma=\frac{2\pi\beta\alpha^{2}}{s}\,g^{2}F_{D}^{2}(s)\left|\psi(0)\right|^{2},\label{eq:sig1}
\end{equation}
where $\beta=k/M_{\Lambda}$~is the baryon velocity, $k=\sqrt{M_{\Lambda}E}$,
$s=\left(2M_{\Lambda}+E\right)^{2}$, $E$~is the kinetic energy
of the pair, $F_{D}(s)=\left(1-s/\Lambda^{2}\right)^{-2}$~is the
dipole form factor, and $\Lambda$~is some parameter close to $\unit[1]{GeV}$.
The factor $g$ is related to the probability of pair production at
small distance $\sim1/\sqrt{s}$ and can be considered as a constant
independent of energy. In Eq.~(\ref{eq:sig1}), $\psi(0)$~is the
wave function of $\Lambda\bar{\Lambda}$ pair at $r=0$. 

The cross section~(\ref{eq:sig1}) is enhanced by the factor $\left|\psi(0)\right|^{2}\gg1$
if there is a loosely bound state or a virtual state of $\Lambda\bar{\Lambda}$
pair. In both cases the modulus of scattering length $a$ of $\Lambda$
and $\bar{\Lambda}$ is large compared to the characteristic radius
$R$ of $\Lambda\bar{\Lambda}$ interaction potential, $\left|a\right|\gg R$.
For a loosely bound state $a$ is positive and the binding energy
is $\varepsilon=-1/M_{\Lambda}a^{2}$. For a virtual state $a$ is
negative and the energy of virtual state is defined as $\varepsilon=1/M_{\Lambda}a^{2}$.
In both cases $\left|\varepsilon\right|$ is much smaller than the
characteristic depth of the potential well. The energy dependence
of $\left|\psi(0)\right|^{2}$ for near-threshold resonances is more
or less universal and is determined by the scattering length $a$
and the effective radius of interaction~\citep{landau3}. Therefore,
one can use any convenient form of potential $U(r)$ for description
of near-threshold resonances.

In the present paper we parametrize the potential as $U(r)=-U_{0}\cdot\theta(R-r)$.
For this potential, the energy dependence of $\left|\psi(0)\right|^{2}$
is well-known (see, e.g., Ref.~\citep{landau3}):
\begin{equation}
\left|\psi(0)\right|^{2}=\frac{q^{2}}{q^{2}\cos^{2}\left(qR\right)+k^{2}\sin^{2}\left(qR\right)}\,,\qquad q=\sqrt{M_{\Lambda}(E+U_{0})}\,.\label{eq:psi}
\end{equation}
Near the threshold $k\ll q$ and the cross section~(\ref{eq:sig1})
is enhanced if
\begin{equation}
q_{0}a\approx\pi\left(n+\frac{1}{2}\right)+\delta\,,\qquad\left|\delta\right|\ll1\,,\label{eq:delta}
\end{equation}
where $q_{0}=\sqrt{M_{\Lambda}U_{0}}$, and $n$~is an integer. For
$\left|\delta\right|\ll1$, the scattering length is $a=1/q_{0}\delta$,
where $\delta>0$ for the bound state, and $\delta<0$ for the virtual
state.

By means of Eq.~(\ref{eq:delta}) the expression~(\ref{eq:psi})
can be simplified:
\begin{align}
 & \left|\psi(0)\right|^{2}\approx\frac{\gamma\,U_{0}}{\left(E+\varepsilon_{0}\right)^{2}+\gamma\,E}\,,\nonumber \\
 & \gamma=4\kappa^{2}U_{0}\,,\qquad\varepsilon_{0}=2\delta\,\kappa\,U_{0}\,,\qquad\kappa=\frac{1}{\pi(n+1/2)}\,.
\end{align}
The corresponding energy dependence of the cross section~(\ref{eq:sig1})
is equivalent to the Flatté formula~\citep{Flatte1976}, which is
expressed in terms of the scattering length and the effective radius~$r_{0}$
of interaction. Note that for the rectangular potential well $r_{0}=R$.
One can easily verify that the precise and approximate formulas for
the cross section are in good agreement with each other for $\left|\delta\right|\ll1$
and $E\lesssim\varepsilon_{0}\ll U_{0}$. Note that $\left|\varepsilon_{0}\right|\gg\left|\varepsilon\right|$
for both bound and virtual states, namely $\varepsilon_{0}\approx2\left|\varepsilon\right|a/R$.
However, the position of peak in the cross section, which is proportional
to $\sqrt{E}\left|\psi(0)\right|^{2}$, is located at energy $E\approx\left|\varepsilon\right|$
for both bound and virtual states.

\begin{figure}
\centering
\includegraphics[totalheight=5.7cm]{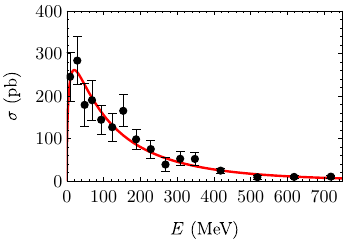}\hfill{}\includegraphics[totalheight=5.7cm]{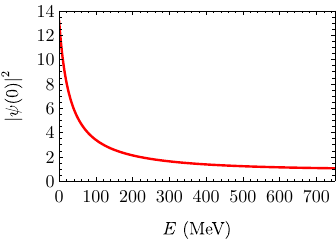}

\caption{The cross section of $e^{+}e^{-}\to\Lambda\bar{\Lambda}$ annihilation
(left) and the enhancement factor $\left|\psi(0)\right|^{2}$ (right)
as the functions of energy $E$. The parameters of the potential are
$U_{0}=\unit[584]{MeV}$ and $R=\unit[0.45]{fm}$. Experimental data
are taken from Ref.~\citep{Ablikim2023}.}\label{fig:plot1}
\end{figure}

In Fig.~\ref{fig:plot1}, we show our predictions for the cross section
compared to experimental data~\citep{Ablikim2023}, as well as the
enhancement factor $\left|\psi(0)\right|^{2}$. The parameters of
the model are $U_{0}=\unit[584]{MeV}$, $R=\unit[0.45]{fm}$, and
$g=0.2$. In the energy region under consideration the dependence
of our predictions on the parameter $\Lambda$ is very weak. To be
specific, we set $\Lambda=\unit[1]{GeV}$. Our model, giving $\chi^{2}/N_{\textrm{df}}=9.8/13$,
provides a good description of experimental data~\citep{Ablikim2023}.
Note that account for the enhancement factor $\left|\psi(0)\right|^{2}$
is of great importance for correct description of experimental data.

Within our model, we also predict a bound state with the binding energy
$E_{0}\approx\unit[-30]{MeV}$. Observation of this bound state would
be very important. The results of Refs.~\citep{Aubert2007c,Chang2009,Ablikim2019b,Ablikim2019d,Ablikim2020a,Ablikim2020b,Ablikim2021g,BESIIICollaboration2021,Xia2022}
indicate the anomalous behavior of the cross sections $e^{+}e^{-}\to K^{+}K^{-}\pi^{+}\pi^{-}$,
$e^{+}e^{-}\to2\left(K^{+}K^{-}\right)$, $e^{+}e^{-}\to\phi K^{+}K^{-}$,
and others at $\sqrt{s}\approx\unit[2.2]{GeV}$ (this value of $s$
corresponds to $E\approx\unit[-30]{MeV}$). However, a more detailed
study of this energy region is required.

In conclusion, the assumption of existence of a low-energy real or
virtual state has allowed us to describe perfectly recent and previous
experimental data for the cross section of $e^{+}e^{-}\to\Lambda\bar{\Lambda}$
annihilation near the threshold. Our model indicates possible existence
of a bound $\Lambda\bar{\Lambda}$ state with energy $E\approx\unit[-30]{MeV}$. 
\begin{acknowledgments}
We are grateful to A.E. Bondar for valuable discussions.
\end{acknowledgments}


\begin{thebibliography}{99}
	\bibitem{Ablikim2023}
	M. Ablikim, et al., \href{https://dx.doi.org/10.1103/PhysRevD.107.072005}{Phys. Rev. D \textbf{107}, 072005 (2023)}.
	\bibitem{Aubert2007}
	B. Aubert, et al., \href{https://dx.doi.org/10.1103/PhysRevD.76.092006}{Phys. Rev. D \textbf{76}, 092006 (2007)}.
	\bibitem{Ablikim2018}
	M. Ablikim, et al., \href{https://dx.doi.org/10.1103/PhysRevD.97.032013}{Phys. Rev. D \textbf{97}, 032013 (2018)}.
	\bibitem{Ablikim2019c}
	M. Ablikim, et al., \href{https://dx.doi.org/10.1103/PhysRevLett.123.122003}{Phys. Rev. Lett. \textbf{123}, 122003 (2019)}.
	\bibitem{Aubert2006}
	B. Aubert, et al., \href{https://dx.doi.org/10.1103/PhysRevD.73.012005}{Phys. Rev. D \textbf{73}, 012005 (2006)}.
	\bibitem{Lees2013}
	J.P. Lees, et al., \href{https://dx.doi.org/10.1103/PhysRevD.87.092005}{Phys. Rev. D \textbf{87}, 092005 (2013)}.
	\bibitem{Akhmetshin2016}
	R.R. Akhmetshin, et al., \href{https://dx.doi.org/10.1016/j.physletb.2016.04.048}{Phys. Lett. B \textbf{759}, 634 (2016)}.
	\bibitem{Ablikim2020}
	M. Ablikim, et al., \href{https://dx.doi.org/10.1103/PhysRevLett.124.042001}{Phys. Rev. Lett. \textbf{124}, 042001 (2020)}.
	\bibitem{Ablikim2021b}
	M. Ablikim, et al., \href{https://dx.doi.org/10.1016/j.physletb.2021.136328}{Phys. Lett. B \textbf{817}, 136328 (2021)}.
	\bibitem{Akhmetshin2019}
	R.R. Akhmetshin, et al., \href{https://dx.doi.org/10.1016/j.physletb.2019.05.032}{Phys. Lett. B \textbf{794}, 64 (2019)}.
	\bibitem{Ablikim2015}
	M. Ablikim, et al., \href{https://dx.doi.org/10.1103/PhysRevD.91.112004}{Phys. Rev. D \textbf{91}, 112004 (2015)}.
	\bibitem{Ablikim2019}
	M. Ablikim, et al., \href{https://dx.doi.org/10.1103/PhysRevD.99.092002}{Phys. Rev. D \textbf{99}, 092002 (2019)}.
	\bibitem{Achasov2014}
	M.N. Achasov, et al., \href{https://dx.doi.org/10.1103/PhysRevD.90.112007}{Phys. Rev. D \textbf{90}, 112007 (2014)}.
	\bibitem{Ablikim2021f}
	M. Ablikim, et al., \href{https://dx.doi.org/10.1038/s41567-021-01345-6}{Nat. Phys. \textbf{17}, 1200 (2021)}.
	\bibitem{Achasov2022}
	M.N. Achasov, et al., \href{https://dx.doi.org/10.1140/epjc/s10052-022-10696-0}{Eur. Phys. J. C \textbf{82}, 761 (2022)}.
	\bibitem{Pakhlova2008}
	G. Pakhlova, et al., \href{https://dx.doi.org/10.1103/PhysRevLett.101.172001}{Phys. Rev. Lett. \textbf{101}, 172001 (2008)}.
	\bibitem{Ablikim2018b}
	M. Ablikim, et al., \href{https://dx.doi.org/10.1103/PhysRevLett.120.132001}{Phys. Rev. Lett. \textbf{120}, 132001 (2018)}.
	\bibitem{Aubert2009}
	B. Aubert, et al., \href{https://dx.doi.org/10.1103/PhysRevLett.102.012001}{Phys. Rev. Lett. \textbf{102}, 012001 (2009)}.
	\bibitem{Haidenbauer2014}
	J. Haidenbauer, X.-W. Kang, and U.-G. Meißner, \href{https://dx.doi.org/10.1016/j.nuclphysa.2014.06.007}{Nucl. Phys. A \textbf{929}, 102 (2014)}.
	\bibitem{Haidenbauer2016}
	J. Haidenbauer and U.-G. Meißner, \href{https://dx.doi.org/10.1016/j.physletb.2016.08.067}{Phys. Lett. B \textbf{761}, 456 (2016)}.
	\bibitem{Haidenbauer2021}
	J. Haidenbauer, U.-G. Meißner, and L.-Y. Dai, \href{https://dx.doi.org/10.1103/PhysRevD.103.014028}{Phys. Rev. D \textbf{103}, 014028 (2021)}.
	\bibitem{Milstein2021}
	A.I. Milstein and S.G. Salnikov, \href{https://dx.doi.org/10.1103/PhysRevD.104.014007}{Phys. Rev. D \textbf{104}, 014007 (2021)}.
	\bibitem{Milstein2022a}
	A.I. Milstein and S.G. Salnikov, \href{https://dx.doi.org/10.1103/PhysRevD.105.074002}{Phys. Rev. D \textbf{105}, 074002 (2022)}.
	\bibitem{Milstein2022}
	A.I. Milstein and S.G. Salnikov, \href{https://dx.doi.org/10.1103/PhysRevD.105.L031501}{Phys. Rev. D \textbf{105}, L031501 (2022)}.
	\bibitem{Milstein2022c}
	A.I. Milstein and S.G. Salnikov, \href{https://dx.doi.org/10.1103/PhysRevD.106.074012}{Phys. Rev. D \textbf{106}, 074012 (2022)}.
	\bibitem{landau3}
	L. D. Landau and E. M. Lifshitz, Quantum Mechanics --- Non-relativistic Theory, Pergamon Press (1991).
	\bibitem{Flatte1976}
	S.M. Flatté, \href{https://dx.doi.org/10.1016/0370-2693(76)90654-7}{Phys. Lett. B \textbf{63}, 224 (1976)}.
	\bibitem{Aubert2007c}
	B. Aubert, et al., \href{https://dx.doi.org/10.1103/PhysRevD.76.012008}{Phys. Rev. D \textbf{76}, 012008 (2007)}.
	\bibitem{Chang2009}
	Y.-W. Chang, et al., \href{https://dx.doi.org/10.1103/PhysRevD.79.052006}{Phys. Rev. D \textbf{79}, 052006 (2009)}.
	\bibitem{Ablikim2019b}
	M. Ablikim, et al., \href{https://dx.doi.org/10.1103/PhysRevD.100.032009}{Phys. Rev. D \textbf{100}, 032009 (2019)}.
	\bibitem{Ablikim2019d}
	M. Ablikim, et al., \href{https://dx.doi.org/10.1103/PhysRevD.99.032001}{Phys. Rev. D \textbf{99}, 032001 (2019)}.
	\bibitem{Ablikim2020a}
	M. Ablikim, et al., \href{https://dx.doi.org/10.1103/PhysRevLett.124.112001}{Phys. Rev. Lett. \textbf{124}, 112001 (2020)}.
	\bibitem{Ablikim2020b}
	M. Ablikim, et al., \href{https://dx.doi.org/10.1103/PhysRevD.102.012008}{Phys. Rev. D \textbf{102}, 012008 (2020)}.
	\bibitem{Ablikim2021g}
	M. Ablikim, et al., \href{https://dx.doi.org/10.1103/PhysRevD.104.032007}{Phys. Rev. D \textbf{104}, 032007 (2021)}.
	\bibitem{BESIIICollaboration2021}
	M. Ablikim, et al., \href{http://arxiv.org/abs/2112.13219}{arXiv:2112.13219 [hep-ex]}.
	\bibitem{Xia2022}
	L. Xia, \href{https://dx.doi.org/10.31349/SuplRevMexFis.3.0308012}{Supl. la Rev. Mex. Física \textbf{3}, 1 (2022)}.
	
\end{thebibliography}
\end{document}